\documentstyle[12pt]{article}
\begin{document}
\baselineskip=17pt
\title{\vspace{-15mm}
        {\normalsize \hfill
       \begin{tabbing}
       \`\begin{tabular}{l}
 	HUB--EP--95/6 \\
        June 1995 \\
        hep--th/9506106 \\
	\end{tabular}
       \end{tabbing} }
       \vspace{10mm}
\setcounter{footnote}{1}
About a class of exact string backgrounds
}

\renewcommand{\thefootnote}{\fnsymbol{footnote}}

\author{~\\
Klaus Behrndt\thanks{e-mail: behrndt@qft2.physik.hu-berlin.de, Work
 supported by the DFG} \\ {\normalsize \em
 Humboldt--Universit\"at, Theorie der Elementarteichen } \\
{\normalsize \em Invalidenstra\ss{}e 110, 10115 Berlin, Germany}
}
\vspace{10mm}
\date{~}
\maketitle
\renewcommand{\arraystretch}{2.0}
\renewcommand{\thefootnote}{\alph{footnote}}
\newcommand{\be}[3]{\begin{equation}  \label{#1#2#3}}
\newcommand{\ee}{ \end{equation}}
\newcommand{\ba}{\begin{array}}
\newcommand{\ea}{\end{array}}
\newcommand{\vsf}{\vspace{5mm}}
\newcommand{\NP}[3]{{\em Nucl. Phys.}{ \bf B#1#2#3}}
\newcommand{\PRD}[2]{{\em Phys. Rev.}{ \bf D#1#2}}
\newcommand{\PRL}[2]{{\em Phys. Rev. Lett.}{ \bf #1#2}}
\newcommand{\MPLA}[1]{{\em Mod. Phys. Lett.}{ \bf A#1}}
\newcommand{\IMP}[1]{{\em Int. J. Mod. Phys.}{ \bf A#1}}
\newcommand{\PL}[3]{{\em Phys. Lett.}{ \bf B#1#2#3}}
\newcommand{\marpar}{\marginpar[!!!]{!!!}}
\renewcommand{\a}{\alpha}
\renewcommand{\b}{\beta}
\newcommand{\g}{\gamma}
\newcommand{\e}{\epsilon}
\newcommand{\f}{\phi}
\newcommand{\m}{\mu}
\newcommand{\n}{\nu}
\renewcommand{\o}{\omega}
\newcommand{\p}{\partial}
\newcommand{\vs}[1]{\vspace{#1mm}}
\newcommand{\s}{\sigma}

\begin{abstract} \noindent
We investigate a class of string backgrounds which have one
conserved chiral null current on the world sheet. In the target space
they have a null Killing vector and unbroken supersymmetries. This
class also known as chiral null model is a generalization of the
gravitational wave and fundamental string background and is exact in
the $\a'$ expansion. The reduction to 4 dimensions yields a stationary
IWP solution which couples to 7 gauge fields (one gravi-photon and 6
matter gauge fields) and 4 scalars. Special cases are the Taub-NUT
geometry and rotating black holes. These solutions possess a
T-self-dual point where the black hole becomes massless.  Discussing
the S-duality we show that the Taub-NUT geometry allows an S-self-dual
point and that the electric black hole corresponds to a magnetic black
hole or an H-monopole. We could identify the massless black hole as
$N_L=0$ and confirm the H-monopole as an $N_L=1$ string states.
\end{abstract}
\thispagestyle{empty}

\newpage

\section{Introduction}
\noindent
String theory is restrictive concerning allowed space time
geometries. The requirement of 2-dimensional (2D) conformal invariance
on the string world sheet fixes allowed backgrounds. There are
mainly two different ways in finding consistent string
backgrounds. One way is to start with a known 2D conformal field
theory and to try to give them a reasonable space time
interpretation. A lot of space time geometries and the corresponding
conformal field theories are known.  However, some classes of
backgrounds seem to be not reachable by this approach, e.g.\ it was
not yet possible to find a reasonable 4-dimensional (4D) black hole or
nonstatic cosmological solution. Fortunately, there is another way as
well. It is known that the conformal points of the world sheet theory
correspond to stationary point of a string effective action which has
directly a space time interpretation. The drawback of this starting
point is that the effective action is known only perturbatively in the
string tension $\a'$. On the other side there are some special classes
of models where the lowest order is already the complete solution,
i.e.\ which do not receive $\a'$ corrections.  These types of
backgrounds are especially interesting if one wants to address
questions about singularities, horizons or the general structure of
the space time. In the last time it was possible to find many
different such solutions including different 4D black holes.  For a
recent review about known exact string solution see \cite{ts2}.
Although they can represent completely different space time geometries
from the string point of view they are often not so different.  Either
one can relate them simply by a dimensional reduction which is nothing
else as a different embedding of the 4D space time into the 10D
target space. Obviously this does not change the 2D world sheet theory.
Or they can be related by string symmetries like T-- or S--duality.
Whereas the T--duality is a symmetry which is valid order by order in
the $\a'$ expansion the situation is not so obvious for the
S--duality, where the different orders in the string loop expansion
get mixed.  This symmetry seems to be of non-perturbative nature and
is up to now not proofed. But let us assume that this conjecture is
right. Thus, a string travelling in such related backgrounds is unable
to distinguish between them and these solutions have to be identified
or in the functional sense one has to sum only once. It is therefore
reasonable to ask what backgrounds can be summarized to equivalent
classes and what are the essential (string) properties of one class.
This is the basic motivation of our investigation.

\vs3

In investigating possible string backgrounds a reasonable assumption
is the existence of at least one Killing vector. On the world sheet
this means the existence of a conserved current. This current becomes
chiral if the metric and the antisymmetric tensor are related in a
certain way. Many exact string backgrounds possess at least one chiral
current, for the WZW model there are even all currents chiral.  If we
further assume that the corresponding Killing vector is null we arrive
at the chiral null model \cite{ho/ts1}.  It is a generalization of the
gravitational wave background and fundamental strings
\cite{ho/ste,kl/ts,da/gi/ha/ru,be/ka/or} and it belongs to the group
of solutions which are already exact in the lowest order in
$\a'$. Furthermore, it has been shown that this model has unbroken
supersymmetries \cite{ka} and thus it is a reasonable bosonic part of
a 10 dimensional (10D) superstring background. The aim of this paper
is to investigate the different solutions belonging to this class.

Since this model gets no $\a'$ corrections it is enough to consider
the lowest order of the effective action only
\be010
\ba{r}
S_{10} = \int d^{10} X \sqrt{\hat{G}} e^{- 2 \hat{\phi}} \left[ \; R +
        4 (\partial \hat{\phi})^2 - \frac{1}{12} \hat{H}^2
          \; \right] + \\
         + \mbox{(higher genus terms)  + (non-pert.\ terms)}
\ea
\ee
where: $X^M = \{v,x^1,...,x^8,u\}$, $\hat{H} = d\hat{B}$ and $\hat{B}$
is the 10D antisymmetric tensor. Of course apart from the $\a'$
expansion there is an expansion in the genus of the world sheet
(string loops) and with respect to this expansion we do not know
much. This expansion goes with the string coupling `constant' ($\sim
g_s = e^{2 \f}$) and the non-perturbative terms are typical of order
${\cal{O}}(\exp{-\frac{1} {2 g_s^2}})$.  We will neglect these
corrections for our further consideration. The starting point is the
10D model and we are going to compactify this theory on a torus
without making any simplifications.  It has already been shown that
after a dimensional reduction one gets electric black holes\footnote{
Strictly speaking as it has been shown in \cite{ka/li} the resulting 4D
objects are not black but white holes which are attracting at long distances
and repulsive for short distances. But for convenience
we will remain at the name black hole.} and Taub-NUT spaces.
This has been done for constant internal space
\cite{ka/ka/or/to} and for the case of an additional modulus field
\cite{ho/ts1}. In a previous paper we have already discussed the
dimensional reduction of the most general case \cite{be}. Here, we are
going to discuss mainly the 4D solutions, the charges, the T-self-dual
point, the S-duality including the S-self-dual point.

\vs3

The paper is organized as follows. In the next section we start
with the gravitational wave and fundamental string background.
The chiral null model is a natural generalization of both. In
section 3 we are going to regard this model as the bosonic part
of a 10D superstring background and will perform the dimensional
reduction to 4 dimensions. As result we find a stationary IWP metric.
In the fourth section we argue that special cases of this metric
are the Taub-NUT geometry and rotating black holes. Also we investigate
the T-self-dual point. Finally, in section 5 we discuss the S-duality
and construct the magnetically charged solutions. At the end we try
to identify special classes of our solutions as elementary string
or solitonic excitations.


\section{Gravitational waves and fundamental strings}

Let us start with two types of string backgrounds which have been
widely discussed in the literature, namely the gravitational wave
background (see e.g., in \cite{ho/ste}) and the fundamental string
background \cite{da/gi/ha/ru}.  After a short
introduction to these models we show that the chiral null model
\cite{ho/ts1} is the natural generalization of both models.

\vs3

If we are talking about gravitational wave background we mean the
pp-waves, which are plane fronted waves with parallel rays.  This
background is defined by an covariantly constant null Killing vector
\be020
D_{M} k_{N} = 0 \qquad , \qquad k_{M} k^{M} =0 ~.
\ee
As solution in general relativity (i.e.~for $\phi=B=0$) this
background is Ricci flat ($R_{MN}=0$) and the metric is given by
\be030
ds^2 = 2 du dv - K(x) du^{2} - dx^I dx^I \qquad , \qquad \partial^2 K(x) =0
\ee
where the harmonic function $K$ in our case depends only on the transversal
coordinates $x^I$; $u$ and $v$ are light cone coordinates. Of course,
this represents a possible string background too, namely for a vanishing
dilaton and antisymmetric tensor. For non-vanishing
antisymmetric tensor we can generalize this wave background to
\be040
\ba{l}
ds^2 = 2 du [ dv -  \frac{1}{2} K(x) du + \o_I(x) dx^I ] - dx^I dx^I \\
B = 2 du \wedge [ dv + \o_I (x) dx^I ]  \qquad , \qquad \hat{\f} = 0 ~.
\ea
\ee
Again the functions $K(x)$ and $\o_I(x)$ should depend on the
transversal coordinates $x^I$ only. The equation of motion for these
functions are discussed below. In principle, one could allow a further
dependence on $u$ (see in \cite{ho/ts1}), but throughout this paper we
assume that all fields are independent of $v$ and $u$.  The covariant
constant vector is given by
\be050
k^{M} = ( k^v=1 , \vec{0}) ~.
\ee
For the special case where $\p_{[I} \o_{J]} = constant$ this model can
be interpreted as a non-semisimple WZW model \cite{ho/ts1} (see also
\cite{na/wi}).

\vs3

Secondly, we are interested in the fundamental string, i.e.~a
macroscopic string in the target space. This background has two
Killing vectors lying in the ``world sheet'' of the this macroscopic
string
\be060
D_{(M} k_{N)} = D_{(M} l_{N)} =0
\ee
and the metric, antisymmetric tensor and dilaton are given by
\be070
\ba{l}
ds^2 = 2 F(x) du [ dv + \o_I dx^I] - dx^I dx^I \\
B = 2 F(x) du \wedge [ dv + \o_I dx^I] \qquad , \qquad e^{2 \hat{\f}} = F(x) ~.
\ea
\ee
For the two Killing vectors one finds
\be080
k^{M} = ( k^v=1 , \vec{0}) \qquad , \qquad l^{M} = ( \vec{0}, l^u =1)~.
\ee
For special choices of the $F$ and vanishing $\o_I$ there exists
again a CFT interpretation as $G/H$ gauged WZW model, where $H$ is
a nilpotent subgroup of $G$ \cite{kl/ts}.

\vs3

Both models can be naturally unified to
\be090
\ba{l}
ds^2 = 2 F(x)\, du [ dv -  \frac{1}{2} K(x) du + \o_I(x) dx^I ] - dx^I dx^I \\
\hat{B} = 2 F(x)\, du \wedge [ dv + \o_I (x) dx^I ]  \qquad ,
\qquad e^{2 \hat{\f}} = F(x) ~.
\ea
\ee
This is the chiral null model which has been discussed by Horowitz and
Tseytlin \cite{ho/ts1}. For $F=1$ we get the gravitational wave
background (\ref{040}) (also known as $K$-model) and for $K=0$ this
model corresponds to the fundamental string (\ref{070})
($F$-model). If one inserts these fields in the 2D world sheet
Lagrangian one finds
\be100
L = 4 F(x) \partial u \, [ \, \bar{\partial} v - \frac{1}{4} K(x)
   \bar{\partial} u +
   \omega_I(x) \bar{\partial} x^I \,] - \partial x^I \bar{\partial} x^I +
   \alpha' R^{(2)} \hat{\phi}(x)~.
\ee
Let us recall some properties of this model. First we see that this
model has a chiral symmetry on the world sheet: $v \rightarrow v +
h(z)$, where $z$ is the complex world sheet coordinate. This
translational invariance of $v$ has in the target space the
consequence that the background admit a Killing vector: $k_{M}$
(\ref{050}) and since there is no kinetic term for $v$ this is a null
Killing vector. Furthermore, since we assume that there is no
dependence in $u$ (generally this is possible) $l_M$ (\ref{080}) is a
Killing vector too.  In the case of gravitational waves ($F=1$)
$k_{M}$ becomes even covariantly constant. In addition, the chiral
symmetry is crucial for the exactness of this model, which means that
all higher $\a'$ corrections in the renormalization group $\beta$
functions vanish. To see this one has to integrate out $u$ and $v$ and
find that for the renormalization only tadpole diagrams are
relevant. The chiral structure of the Lagrangian makes it impossible
to construct other (non-tadpole) divergent diagrams (for details see
especially the second Ref.~in \cite{ho/ts1}).  Thus, the conformal
invariance conditions are given by the lowest order in $\alpha'$ and,
if we drop a linear dilaton part, we have the equations
\be110
-\partial^2 K(x) \, = \, -\partial^2 F^{-1}(x) \, = \, 0
\quad , \quad -\partial^I F_{IJ} \, = \, 0 \qquad \mbox{and} \qquad
 e^{2 \hat{\phi}} \sim F(x) \ .
\ee
($F_{IJ}=\partial_{[I} \omega_{J]}$). These are the equation of motion
for the fields in (\ref{100}).  Investigating the T-duality we find
that only the (harmonic) scalar functions $K$ and $F^{-1}$ are
mixing. If we take an arbitrary direction in the $(u,v)$ plane, e.g.\
by transforming $v = \hat{v} + c \,u$, then dualizing $u$ and finally
reverse the $v$ shift we find
\be120
F' = \left(2 c - K\right)^{-1} \quad , \quad K' = 2c - F^{-1}  \quad ,
\quad \omega'_I = \omega_I
\quad , \quad e^{-2 \hat{\phi}'} = F'^{-1}
\ee
Using the fact that $K$ and $F^{-1}\sim e^{-2 \hat{\phi}}$ are harmonic
functions we see that the duality transformation changes only
the parameter of the solution. The model is explicitly self-dual,
i.e.\ the functions $F$ and $K$ remain unchanged if
\be130
K + F^{-1} = 2 c\ .
\ee
Fixing the asymptotic values to $K|_{\infty}=F|_{\infty}=1$,
we see that the model is explicitly self-dual only for $c=1$, i.e.\ not
for arbitrary directions in the $(u,v)$ plane. Obviously, the
wave background (\ref{040}) with $F=1$ is dual to the fundamental
string (\ref{070}) with $K=0$ and only if both scalar fields are
nontrivial we can have self-duality.

Finally, let us note that embedded in N=1, D=10 supergravity this
model admits unbroken supersymmetries \cite{ka}, \cite{da/gi/ha/ru},
\cite{ka/ka/or/to}, \cite{wa}. On the other side one has to note that for
the exactness of this model the supersymmetry is not crucial.
Instead, as it has been shown in \cite{ho/ts1} the pure bosonic
background is already exact in all orders in the $\a'$ expansion.

\section{The IWP solution in 4 dimensions}
Of course, it is possible to regard the model (\ref{100}) directly in
4 dimensions. But instead, the unbroken supersymmetries suggest to
consider this model as the bosonic part of a 10D superstring
background. Then we can ask, what does the corresponding 4D theory
look like? Following this question we start with the 10D effective
action, perform the dimensional reduction and discuss the 4D fields.
In the next section we will relate the results to Taub-Nut and
rotating black hole solutions.

\vs3

Since for our model (\ref{100}) there are no higher $\alpha'$
corrections the complete effective action (up to non-perturbative and
higher genus contributions) is given by (\ref{010}). Let us now start
with the dimensional reduction.  Since the reduction procedure
preserves the supersymmetry also the 4D background has unbroken
supersymmetries (corresponding to N=4).  Assuming that the theory does
not depend on 6 coordinates and that the internal space is compact we
can integrate over internal coordinates and get the 4D theory. This is
more or less standard in string compactification (see e.g.\ in
\cite{ma/sc}). On the other side if one admits a dependence on the
internal coordinates, one can make a Fourier expansion in the internal
coordinates. After reduction one gets then states with masses
corresponding to the inverse compactification scales (see
e.g.~\cite{du}). In this philosophy we are in the massless sector.

Now, we have to embed the 4D space time into the 10D target space.
We choose $v$ as time coordinate and three of the transversal coordinates
as spatial part, i.e.\ $x^M=(v , x^i | x^r , u)$ and get for
the metric and antisymmetric tensor (\ref{090})
\be140
\hat{G}_{MN} =  \left( \ba{cc|cc}
0 & 0 & 0  & F \\[-4mm]
0 & -\delta_{ij} & 0 & F \,\omega_i \\ \hline
0 & 0 & -\delta_{mn} & F\,\omega_m  \\
F & F\,\omega_i  &F\,\omega_m  & - F\,K \ea \right) \quad , \quad
\hat{B}_{MN} =  \left( \ba{cc|cc}
0 & 0 & 0  & F \\[-4mm]
0 & 0 & 0 & F\,\omega_i \\ \hline
0 & 0 & 0 & F\,\omega_m  \\
-F & -F\,\omega_i  &-F\,\omega_m  & 0 \ea \right)
\ee
where the last line corresponds to the $u$ coordinate. It seems to be a
little disturbing that the former light cone coordinate $u$ becomes
now part of the internal space. But as long as we make sure that the
internal space remains compact and the 4D space time metric has the
right signature there is no reason against this.  Following now the
standard procedure for dimensional reduction (see e.g.\
\cite{ma/sc,ch}) we write the 10-Bein as
\be150
e_M^{\ \hat{N}} = \left( \ba{c|c}
        e_{\mu}^{\ \hat{\mu}} & A_{\mu}^{\ r} E_r^{\ \hat{s}} \\ \hline
         0         &   E_r^{\ \hat{s}} \ea \right) \ .
\ee
The 4D space-time metric is given by $g_{\mu\nu} = e_{\mu}^{\hat{\mu}}
e_{\nu}^{\hat{\nu}} \eta_{\hat{\mu} \hat{\nu}}$ and the internal
metric is $G_{rs} = E_{r}^{\hat{r}} E_{s }^{\hat{s }} \delta_{\hat{r }
\hat{s }}$ ($r,s = 4,..  9$).  This form of the 10-Bein has the
advantage that the determinant of the metric and thus the volume
measure factorizes and we can absorb the internal part in the dilaton
\be160
\sqrt{|\hat{G}_{MN}|} \, e^{-2 \hat{\phi}} = |e_M^{\ N}|
\, e^{-2 \hat{\phi}} =
\sqrt{|g_{\mu\nu}|} \sqrt{|G_{rs}|} \, e^{-2 \hat{\phi}} =
\sqrt{|g_{\mu\nu}|} e^{-2 \phi}
\ee
with the 4D dilaton $\phi$ defined by
\be170
\sqrt{|G_{rs}|} \, e^{-2 \hat{\phi}} = e^{-2 \phi}  \ .
\ee
Using the 10-Bein (\ref{150}) we can write the 10D metric as
\be180
\hat{G}_{MN} = \left( \ba{c|c}
  g_{\mu\nu} + A^{~r}_{\mu} A_{\nu r} &
  A_{\mu r}\\ \hline
  A_{\mu s} & G_{rs} \ea \right) \ .
\ee
Comparing (\ref{180}) with (\ref{140}) we find for the internal metric
and the gauge field part
\be190
G_{rs} =  \left(\ba{c|c} -  \delta_{mn} & F\,\omega_m \\[2mm] \hline
           F\,\omega_n  & - F K \ea \right)  \qquad , \qquad A_{\mu r} =
       \big( \  0 \  | \, A_{\mu} \, \big)
\ee
with $A_{\mu} = F\, \{1, \omega_i \} $ and  inserting
this  metric in (\ref{170}) yields for the 4D dilaton
\be200
 e^{-2 \phi} =  e^{-2 \hat{\phi}}  \sqrt{F K - F^2\,|\omega_m|^2} =
\sqrt{F^{-1} K - |\omega_m|^2}
\ee
where $|\omega_m|^2 = \omega_4^2 + \omega_5^2 + .. $ . In the
following we are assuming that
\be201
F^{-1} K > |\omega_m|^2\ .
\ee
This condition means that all eigenvalues of the internal metric are
negative\footnote{The eigenvalues of $G_{rs}$ are:
$\left\{-1,-1,-1,-1,-\frac{1}{2}\left((1 + F K) \pm \sqrt{(1 - F K)^2
+ 4 F^2 |\omega_m|^2}\right)\right\}$.} , i.e.\ there is no timelike
compactified coordinate. Therefore, this condition ensures that the
internal space remains compact, i.e.\ even $u$ represents a spacial
direction. In discussing our results we will return to this point
once more (after eq.\ (\ref{306})) but here let us proceed in deriving
the 4D fields. For the metric we get finally
\be210
ds^2 = \frac{1}{F^{-1} K - |\omega_m|^2} (dt + \omega_i dx^i)^2 -
  dx^i dx^i =  e^{4 \phi} (dt + \omega_i dx^i)^2 - dx^i dx^i .
\ee
For the 4D antisymmetric tensor we are using the definition of
\cite{ma/sc} and find that it vanishes in our case
\be220
B_{\mu\nu} = \hat{B}_{\mu\nu} - A_{\mu}^{\ r} \hat{B}_{rs}
 A_{\nu}^{\ s} = 0
\ee
(where $\hat{B}_{\mu\nu}$ are the 10D components of (\ref{090})).  In
principle there is one further term which, however, vanishes too,
because the off-diagonal terms coming from the metric and
antisymmetric tensor are equal. But nevertheless the 4D torsion is
non-vanishing and is proportional to the Chern-Simons term of the
gauge field $A_{\mu}=F \{1,\omega^i\}$
\be230
H_{\mu\nu\rho} = - A_{[\mu}^{\ \ r} \partial_{\nu} A_{\rho] s} =
         - e^{4 \phi}\, F^{-2}\, A_{[\mu} \partial_{\nu} A_{\rho]} \ .
\ee
Using the equation of motion for $\o^i$ we can write the torsion
on shell in a different form. From (\ref{110}) we get
\be270
\p ^i \partial_{[i} \omega_{j]} =  0 \ .
\ee
Immediately, we find the two solutions
\be280
\partial_{[i} \omega_{j]} = \epsilon_{ijk} \partial^k \, a(x) \qquad
\mbox{or} \qquad \partial_{[i} \omega_{j]} = const.
\ee
In the second case we have an uniform magnetic field and under certain
assumptions the target space is parallelizable and the model corresponds
to a product of a non-semisimple WZW model and a free spatial
direction. The corresponding 4-D space time is not asymptotically
flat.  Let us ignore this case here (see \cite{ru/ts}).  The first
case defines a further harmonic scalar field $a$. Inserting
this expression into (\ref{230}) we find for the torsion
\be290
(\det g)^{- \frac{1}{2}} \epsilon^{\lambda\mu\nu\rho} H_{\mu\nu\rho} =
    - e^{2 \phi} \, \partial^{\lambda} \, a
\ee
Therefore, $a$ is the axion field which determines the torsion ($H =
e^{4 \phi}\, ^*a$, in the Einstein frame).

Next, we have to discuss the gauge fields. Gauge fields
appear in the Kaluza--Klein procedure as non-diagonal components of
the metric and the antisymmetric tensor. The gauge fields coming from
the metric are in principle given by (\ref{190}). But there is a
subtlety.  Investigating the gauge transformation one realizes
that the basic gauge fields have an upper internal index. The
reason is, that gauge transformations are generated by local
translations in the internal coordinates which have an upper index.
Rising the index we get
\be250
A_{\mu}^{\ r} = A_{\mu s} G^{sr} = - \frac{1}{2} \, e^{-2 \s}
(F \omega^r + l^r) A_{\mu} \quad , \quad \o^r = ( \o^m , 0) \quad ,
\quad l^r = (\vec{0},1)
\ee
($\omega^m = \omega_m$, $l^r$ is the second Killing vector of the 10D
theory). On the other side, the gauge fields coming from the
antisymmetric tensor have a lower internal index. The corresponding
gauge transformation is part of the antisymmetric tensor gauge
symmetry. But here, to get the right gauge field coupling in the
effective action we have to add an additional term \cite{ma/sc} and
obtain
\be260
B_{r \mu} = \hat{B}_{r \mu} - \hat{B}_{rs} A_{\mu}^{\ s} =
\frac{1}{2}\, e^{-2 \s} (F \omega^r - K F \;l^r) A_{\mu} \ .
\ee

Finally, in the scalar field sector we have the dilaton, the axion and
two modulus fields. The 4D dilaton is given by (\ref{200}) and the
axion is a not yet specified harmonic function. That there are at
least two moduli fields becomes already obvious if we look at the
eigenvalues of the internal metric (see footnote on page 8).  If we
diagonalize the internal metric we find that the internal space
factorizes in a trivial 4D space (with vanishing antisymmetric tensor)
and a non-trivial 2D space with the antisymmetric tensor and metric
given by
\be261
G_{mn} = \left( \ba{cc} -\lambda_1 & 0 \\
0  & - \lambda_2 \ea \right) \quad , \quad
B_{mn} =  \left( \ba{cc}
0 & F\,|\omega_m|  \\ -F\,|\omega_m|  & 0 \ea \right)
\ee
where $\lambda_{1/2}$ are the eigenvalues of the internal metric given
in the footnote on page 8. We see that this 2D space is determined
by two scalar functions
\be262
e^{\sigma} = \sqrt{G_{mn}} = \sqrt{F K - F^2 |\omega_m|^2} \qquad , \qquad
\eta = \sqrt{B_{mn}} = F |\omega_m|
\ee
which we can combine to a complex scalar field
\be263
T = \eta + i e^{\sigma} \ .
\ee
This complex scalar function $T$ parameterizes the non-trivial 2-torus.
Note, that the T-duality transformation given in (\ref{120}) with the
self-dual
point defined in (\ref{130}) is not taken with respect to this $T$ modulus
field. The crucial point in our duality transformation was that we first
shifted the time $v$, then performed the duality transformation and finally
inverted the time shift, i.e.\ we have
``given the isometry direction $u$ a piece of time''.

\vs3

\noindent
Summarizing our results the general 4-D solution is given by
\be300
\ba{ccc}
ds^2 = e^{4 \phi} (dt + \omega_i dx^i)^2 - dx^i dx^i & , &
 e^{-2 \phi} = \sqrt{K F^{-1}  - |\omega_m|^2} \\
 H^{\mu\nu\rho} = - e^{2 \phi}\,(\det g)^{-\frac{1}{2}}\,
\epsilon^{\mu\nu\rho\lambda}  \partial_{\lambda} a  & , &
 e^{2 \sigma} = e^{4 ( \hat{\phi} - \phi )} =  K\, F  - F^2\,|\omega_m|^2
\ea
\ee
with $\omega_m = \omega^m$. The metric is the string version of the
Israel-Wilson-Perjes (IWP) metric. In Einstein-Maxwell gravity this
metric was constructed as a general stationary solution
\cite{is/wi/pe}.  Basically, this metric has a non-flat spacial part,
but Tod \cite{to} found that the class of stationary metrics admitting
a Killing spinor is given by the IWP metrics with flat 3D space.
This class can then be embedded into supergravity.  The static limit
($\o_i=0$) is given by the Majumdar--Papapetrou black hole.
Well-known examples for IWP metrics are Taub-NUT and rotating black
hole space times. The gauge fields can be combined as following
\be302
\left( \ba{c} A_{\m}^{(+) r}\\ A_{\m}^{(-) r}\ea\right) = \frac{1}{\sqrt{2}}
\left(\ba{l} A_{\m}^{\ r} + B_{r \m} \\
             A_{\m}^{\ r} - B_{r \m} \ea \right) = -\,\frac{1}{\sqrt{2}}
\,e^{-2 \s} \left(\ba{l} \frac{1}{2} (K F + 1)\, l^r  \\
       F \,\o^r - \frac{1}{2} (K F - 1)\,l^r \ea \right) A_{\m} \ .
\ee
with $ A_{\mu} = F\, \{\,1\,,\,\omega_i\}$. Thus, after the
dimensional reduction we have at all three scalars (dilaton, moduli
field and axion) and 7 vectors. Because we have started with a
supersymmetric theory in 10 dimension we expect to arrive in a N=4
theory in 4 dimensions. Indeed, the vector fields decompose into 6
matter vectors ($A_{\m}^{(-)r}$) and one gravi-photon ($A_{\m}^{(+)r}$)
\cite{ch}.

All these fields are completely determined by eight harmonic functions
\be301
-\partial^2 K(x) \, = \, -\partial^2 F^{-1}(x) \, = -\partial^2 \omega_m(x)
\, = \,- \partial^2 a(x) \, = \, 0
\ee
where the derivatives acts only on the three spacial coordinates $\vec{x}$.
Note, that the vector $\o ^i$ is fixed by (\ref{280}) and $m$ is an internal
index so that all functions are space time scalars.

Before we further investigate the general solution let us discuss the
relation to other known solutions. First, if we take the internal
space flat, i.e.\ $\omega_m =0$ and $ F K =1$. We arrive at the IWP
solution discussed in \cite{ka/ka/or/to}. In this case all matter
gauge fields vanish ($A_{\m}^{(-)r}=0$) and in addition we have no
modulus field ($\sigma=0$). Second, Horowitz and Tseytlin
\cite{ho/ts1} have discussed the case $\omega_m =0$ but $F K \not=1$,
i.e.\ they have added a nontrivial modulus field and one further
(matter) gauge field.  One example is the Kaluza--Klein (KK)
solution. Setting $\o_i = \o_m =0$ and $F=1$ in the 10D theory the
antisymmetric tensor and dilaton vanish and we arrive at the
Einstein-Hilbert action. The solution is just the gravitational wave
background (\ref{030}) and after the reduction we have no torsion and
only one gauge field coming from (\ref{250}). Inserting a harmonic
function for $K$ this solution in the Einstein frame is given by the
known extremely charged KK-black hole \cite{gi}
\be306
\ba{cc}
ds^2 = e^{2 \phi} dt ^2 - e^{-2 \phi} dx^i dx^i & , \quad
 e^{-4 \phi} =  e^{2 \sigma} = 1 + \frac{2 m}{r}  \\
 H_{\m\n\rho} = 0 &, \quad A_{0} = - \frac{1}{2} \,
 (1 + \frac{2 m}{r})^{-1}  \ .
\ea
\ee

Let us now return once more to the condition (\ref{201}).  This
condition was crucial for getting an euclidean internal space. If this
condition is not fulfilled the internal metric has a positive
eigenvalue (see footnote on page 8) corresponding to one timelike
internal direction and therefore a non-compact internal space.  On the
other side for the fundamental string solution we have $K=0$, and
thus, this condition is not valid.  This becomes clear if one
remembers how to construct the fundamental string solution
\cite{ho/st}. One can start with a (uncharged) black string as a
direct product of the Schwarzschild metric and a flat direction. After
a Lorentz boost in the flat direction one can dualize this direction
and get a dilaton and an $H$-charge. Finally, after performing the
extremal limit one gets the fundamental string winding around the flat
direction. The coordinates $u$ and $v$ are there light cone
coordinates, which are both non compact. There are two possibilities
to avoid this problem. First is to take at least one $\omega_m$
imaginary corresponding to a Wick rotation in the internal coordinate
with the wrong eigenvalue. This was done in \cite{be/ka/or} to get a
black hole solution from the fundamental string in 10
dimensions. Another way is to give the fundamental string first a
non-zero linear momentum and reduce it then. In this case $K\not=0$
and can be normalized to $K=1$ \cite{ho/ts1,wa}. In both cases one can
find a region with right signature.

\vs3

Before we turn to special examples we want to discuss the theory in
the neighborhood of the singular points of the scalar functions $F$,
$K$, $a$, $\o^r$ (\ref{301}). At these points our metric and dilaton
as well as the other 4D fields are singular. Although our solution is
exact in the $\a'$ expansion we should expect that at these points higher
genus (string loop) contributions and non-perturbative terms become
important. But this not the case.  If we remember that the string
coupling constant $g_s$ is defined by the dilaton we find near the
singularities
\be303 g_s^2 = e^{4 \f} = \frac{1}{K F^{-1} - |\o^r|^2}
\longrightarrow 0 \ ,
\ee
i.e.\ the string coupling constant vanishes and therefore the higher
genus contributions ($\sim g_s$) and the non-perturbative terms ($\sim
\exp\{- \frac{2}{g_s^2}\}$) are expected to be under control.  Thus
our model does not break down there but seems to behave even better,
it becomes asymptotically free. This feature has been discussed for
the wave or fundamental string background in \cite{ts}.

\section{Relation to Taub-NUT and rotating black hole solutions}

In the last section we have performed the dimensional reduction of our
model and have discussed the field content in 4 dimensions.  The 4D
solution (\ref{300}) is determined by eight harmonic functions
$K(x)$, $F^{-1}(x)$, $a(x)$, $\o^r(x)$. In this section we are going
to discuss special examples for these functions and we will find
different space time geometries.

\vs3

The general solution for the harmonic functions is given by the real
or imaginary part of
\be305
\sim c_0 + \sum_{k=1}^{N} \frac{c_k}{r_k}
\ee
where $r_k^2 = (x-x_k)^2 + (y-y_k)^2 + (z-z_k)^2$ and the parameter
($c_0, c_k , x_k, y_k, x_k$) are complex.  Now we have to choose these
complex parameter in a proper manner.  Calculating the real and
imaginary part one finds that the physical distinguishable situations
are determined by the different singularity structures. That means,
without loss in generality we can assume $c_0$ and $c_k$ as
real. Furthermore, to make the situation more clear we assume that we
have only one center, i.e.\ no sum over $k$
($\vec{x_k}\rightarrow\vec{x}_0$).  In this case we can absorb the
real part of $\vec{x}_0$ by shifting the spatial coordinate $\vec{x}$,
and we set $\vec{x_0}=i(\g,\b,\a)$ ($\a,\b,\g$ are now real
parameter).  Thus, we write the harmonic functions as
\be304
\sim c_1 + \frac{c_2}{\sqrt{(x-i\g)^2 + (y - i\b)^2 + (z - i\a)^2}}
\ee
We have now to distinguish that {\em all, two, one} or {\em none}
of the imaginary parts are zero. Understanding this situation
the generalization to multi-center solutions is straightforward.

\vs5

\noindent
{\em 1. Taub-NUT: $\a=\b=\g=0$.} Here, the imaginary parts vanish and
the theory is completely spherical symmetric. We can write the harmonic
functions as
\be310
K = 1 + \frac{2 m}{r} \quad , \quad F^{-1} = 1 + \frac{2 \tilde{m}}{r}
\quad , \quad \omega_m = \frac{2 q^m}{r} \quad , \quad a = \frac{2 n}{r}
\ee
where $r^2 = x^2 + y^2 + z^2$ and $\vec{\o}$ is defined by the axion
$a$ ($\p_{[i} \o_{j]} = \e_{ijk} \p_{k} a$).  In this case the
harmonic functions $K$ and $F^{-1}$ can always be written as $K = 2 c
- b F^{-1}$. The constants $c$ and $b$ can be adjusted by a proper
choice of the $u$ and $v$ coordinates in 10 dimensions (translation in
$v$ and scaling of $u$) and one finds that only $b=\pm 1$
($m=\pm\tilde{m}$) are non-trivial. We will often distinguish between
these two cases.  The self-dual case (\ref{130}) is given by $b=1$ or
$m=-\tilde{m}$ (to get the right asymptotic behavior we set $c=1$).
The other case ($m=\tilde{m}$) is crucial for a flat internal
space. Inserting these fields and a solution for $\omega_i$
in (\ref{300}) yields for the metric and dilaton
\be340
ds^2 = e^{4 \phi} \left( dt + 2 n \cos \theta d\phi \right)^2 -
\left( dr^2 + r^2 d\Omega^2 \right) \quad , \quad
e^{4 \phi} = \frac{1}{(1 + \frac{r_+}{r})(1 + \frac{r_-}{r})}
\ee
with
\be350
r_{\pm} = m + \tilde{m}  \pm \sqrt{(m-\tilde{m})^2 + 4 |q_m|^2} =
\left\{ \ba{ccc} \pm 2 \sqrt{ \tilde{m}^2 + |q|^2} & , & m=-\tilde{m}\\
2(m \pm |q|) & , & \ m=\tilde{m} \ea \right. \ .
\ee
This is an extremely charged Taub-NUT solution. If we make our
internal space flat ($\o_m =0$, $FK=1$ or $r_+ = r_-$, $m=\tilde{m}$)
this solution again coincides with solution found in
\cite{ka/ka/or/to}, \cite{jo/my}. This metric has a wire singularity,
which makes it impossible to invert the metric along the axes
$\theta=0,\ \pi$. To see this one can consider slides with
constant $t$ and $r$ for which $ds^2$ does not vanish at the north and
south pole (as one would expect for a smooth space time).  Though this
singularity can be removed by choosing different times at the north
and south hemisphere ($t \rightarrow t \pm 4\pi \f$).  But for consistency
at the overlapping region one has to require that the time is periodical
$t \sim t + 8 \pi n$ (see e.g.\ in \cite{mi}). Also, this metric is
asymptotically not flat.  Instead, one can show that for constant
radius this metric has an $S_3$ geometry.

Furthermore, if $r_-<0$ ($m\tilde{m} < |q_m|^2$) the metric contains a
pole at $r=r_-$. An example is the self-dual case (\ref{130}) where
$r_-=-r_+$ ($m=-\tilde{m}$).  What happens at this point?  If we
consider the corresponding moduli field $e^{2 \sigma}$ (\ref{262})
\be360
e^{2 \sigma} = \lambda_1 \lambda_2 = \frac{r+r_+}{r+2\tilde{m}}
 \; \frac{r+r_-}{r+2\tilde{m}} \quad \longrightarrow \quad \left\{
\ba{ll} 1 \quad & , \quad \mbox{ for } r \rightarrow \infty \\
    \frac{m \tilde{m} - |q_m|^2}{\tilde{m}^2} \quad & , \quad \mbox{ for }
    r \rightarrow 0 \ea \right.
\ee
where $\lambda_{1/2}$ are the eigenvalues of the internal metric which
define the radii of the non-trivial torus.  We see that for negative
$r_-$ one radius has a zero and becomes negative for $r < -r_-$,
and thus, this coordinate becomes time like. Also at this point the
space time metric changes its signature and becomes euclidean.
It has been shown in \cite{ka2} that at this point the space time has
a  true singularity. Interestingly, the 10D theory is  here
completely smooth. In some sense the singularity is compensated
by the vanishing volume of the internal space ($\det{G_{rs}}=0$).
As it was shown by Kallosh and Linde \cite{ka/li} this singularity
becomes smooth and non-singular on the quantum level.
On the other side for positive $r_-$ the compactification
radii are bounded for all $r$, and thus, the internal space remains
``invisible'' (as long as we choose the compactification scale
sufficiently small).

\vs3

\noindent
{\em 2. Rotating black holes: $\a\neq0$, $\b=\g=0$.} Here the theory
is axial-symmetric and the singularities of the scalar functions are
in the plane $z=0$ on the circle $x^2 + y^2 = \a^2$. We write
\be361
\ba{ccc}
K = Re \left( 1 + \frac{ 2 m}{\sqrt{x^2 + y^2 + (z-i\a)^2}}\right)
& , &
F^{-1} =  Re \left( 1 + \frac{ 2 \tilde{m}}{\sqrt{x^2 + y^2 +
(z-i\a)^2}}\right) \\
\o^r = Re \left( \frac{ 2 q^r}{\sqrt{x^2 + y^2 + (z-i\a)^2}}\right)
& , &
a =  Im \left( \frac{ 2 n}{\sqrt{x^2 + y^2 + (z-i\a)^2}}\right) \ .
\ea
\ee
As we will see below this choice corresponds to the inclusion of
non-vanishing angular momentum which is proportional to $\a$.  In
order to get back the static solution\footnote{In the classification
of time independent metrics static means, that one can diagonalize the
metric. In contrast to stationary metrics, which are also time
independent, but the non-diagonal part corresponds, e.g., to a
non-vanishing angular momentum or Taub-NUT charge.} ($\vec{\omega}=0$)
for vanishing angular momentum ($\a =0$) we have to take the imaginary
part for the axion field and the real part for the other functions.
So, taking these functions and transforming the solution
into spheroidal coordinates ($x+iy = \sqrt{r^2 + \alpha^2} \sin
\theta \exp\{\pm i \phi\}$; $z = r \cos \theta$) we find
\be370
\omega_{\phi} = \frac{2 n \a \sin^2 \theta}{R} \qquad \mbox{with}
\qquad R = \frac{r^2 + \a^2 \cos^2 \theta}{r}
\ee
and for the metric and the dilaton
\be380
\ba{l}
ds^2 =e^{4 \phi}
\left( dt + \omega_{\phi} d\phi \right)^2 - d\vec{x}^2 \qquad , \qquad
e^{4 \phi} =  \frac{1}{(1 + \frac{r_+}{R})(1 + \frac{r_-}{R})} \\
 d\vec{x}^2 =
\frac{r^2 + \a^2 \cos^2 \theta}{r^2 + \a^2} dr^2 +
(r^2  + \a^2 \cos^2 \theta)d\theta^2 +(r^2 + \a^2)
 \sin^2\theta d\phi^2
\ea
\ee
where $R=0$ defines the ring singularity. Here, as for the Taub-NUT
type solution (\ref{340}) the only influence of the additional gauge
fields and the moduli field is a splitting in the dilaton field ($r_+
\not= r_-$). Again for a flat internal space this solution coincides
with \cite{ka/ka/or/to}.

In difference to the Taub-NUT case this solution is asymptotically
flat and we can define conserved charges. First from the asymptotic
behaviour of metric in the Einstein frame we get the  mass (from
$g_{00}^{E}$) and the angular momentum (from $g_{0\f}^{E}$)
\be381
M= \frac{1}{4} (r_+ + r_-) = \frac{1}{2} (m + \tilde{m}) \qquad , \qquad
J= n\, \a
\ee
and from the gauge fields we can read off the electric charges
($\vec{A}_0 \propto - \frac{\vec{Q}_e}{r}$) and find
\be382
\left( \ba{c} Q_e^{(+)r} \\ Q_e^{(-)r} \ea \right) =
\sqrt{2}\left( \ba{c} \frac{1}{2} (m+\tilde{m}) \, l^r \\
- q^r - \frac{1}{2} (m-\tilde{m}) \, l^r \ea \right) \qquad , \quad
q^r = ( q^m , 0 ) \ .
\ee
Now we can express the constants $r_{\pm}$ in the solutions by these
charges and obtain
\be383
r_{\pm} = \sqrt{2} \left(\, |\vec{Q}^{(+)}| \pm  |\vec{Q}^{(-)}|\, \right)
\ee
Investigating the dual field strength tensor we see that there are
no magnetic charges but magnetic moments ($\tilde{F}_{0r} \propto 2
\frac{\m}{r^3} \, \cos \theta $)
\be390
\m^{(-)r} = 0  \quad , \quad \m^{(+)r} =
 \sqrt{2}\, n\,\a \, l^r \ .
\ee
Having the charges we can ask which of the states saturate a
Bogomol'nyi bound. Following the procedure of Sen \cite{se2}
we get
\be391
M^2 = \frac{1}{2} \,e^{-2\f_0} \, \vec{Q}_e \cdot (L M_0 L + L)
\cdot \vec{Q}_e  = \frac{|\vec{Q}^{(+)}|^2}{2}
\ee
(the index ``$0$'' means asymptotical values and $(L M_0 L + L)$ is in
our case just the projector on the $(+)$ states). Thus, as expected
for a supersymmetric solution the Bogomol'nyi bound is saturated.  But
only the gravi photon sector ($A_{\m}^{(+)r}$) enter this bound and
the matter gauge fields sector ($A_{\m}^{(-)r}$) are not restricted.
Therefore, the gravi photon has to be a lowest state.

An interesting case in this context is the self-dual configuration
(\ref{130}) where $\tilde{m} = - m$. Here the black hole states
become massless ($M=0$) and carry only the matter gauge field
charges $\vec{Q}^{(-)}$ ($\vec{Q}^{(+)}=0$). This is consistent
with the statement that self-dual configurations are points of
enhanced symmetry and often correlated with additional massless
states. But note that our duality transformation is not taken with
respect to the internal space only but a mixing of the time and
the internal coordinate $u$. The self-dual point with respect to the
$T$ modulus field (e.g.\ $T=i$) does not  correspond to massless
black holes.

Let us end this section with a discussion of the singularity.  As it
was pointed out in \cite{ka/ka/or/to} the solution with flat internal
space exhibits a naked singularity.  Since the causal structure and
the singularities are not affected by a non-flat internal space (as
long as $r_- > 0$) the metric (\ref{380}) has still a naked
singularity.  This can also be seen by a direct comparison with a
rotating black hole solution. Starting with a 4-D Kerr solution and
using the O(d,d+p) technique Sen \cite{se} has constructed a general
black hole solution including 28 U(1) gauge fields and nontrivial
moduli. One can show that both solutions coincide in a special
limit\footnote{We have to set $m=\tilde{m}= \sqrt{n^2 + |q|^2}$ in our
solution and take in Sen's solution the limit $m^{Sen} \rightarrow 0$,
$\b \rightarrow \infty$, but $m \, \sinh \b \, \cosh \a = 2 M$ remains
fix ($|q_m|=M^{Sen} \tanh \a^{Sen}$, $a^{Sen} = \a \cosh
\a^{Sen}$). Note, that in Sen's solution all parameters are correlated
to each other.  But in our solution, e.g., $n$ is not restricted. The
reason is that Sen has generated his solution by a $O(d,d)$
transformation, which does not yield the most general rotating black hole
solution.}. On the other side, Sen's solution has two horizons, which
vanish in this limit and the singularity becomes naked.  But what is
the interpretation of this singularity? There has been a lot of
discussion that the ring singularity of a Kerr black hole can be
regarded as closed elementary string. E.g. Nishino \cite{ni} showed
some indication that the Kerr solution with a naked singularity is
nothing else as the gravitational field generated by a closed
string. Furthermore, this interpretation could yield for the entropy
the correct density of elementary string states \cite{se3}. Let us
discuss here only one feature which supports this idea, namely the
gyromagnetic ratios.  They can be defined by $\vec{\m} = \frac{1}{2}
\, g \, J\, \frac{\vec{Q}}{M}$ and for our solution we obtain
\be400
g^{(-)} =0 \quad , \quad g^{(+)} = 2
\ee
i.e.\ the matter gauge fields have a zero gyromagnetic ratio and the
gravi-photon has $g=2$. As pointed out by Sen \cite{se} these values
coincides with the corresponding values for elementary string
states. Using a generalization of the results of Russo
and Susskind \cite{ru/su} Sen found for $g$
\be401
g^{(\pm)} = 2 \frac{S^{(\mp)}}{S^{(-)} + S^{(+)}}
\ee
where $S^{(\pm)}$ is the contribution to the $z$ component to the
angular momentum. Since in our case the fields in the right moving
sector ($+$ component) saturate the Bogomol'nyi bound they have to be
in the lowest state. Therefore the angular momentum coming from this
part can be neglected with respect to left moving part and one finds
the values (\ref{400}). In addition the value $g=2$ for the
Bogomol'nyi states is just the natural value of $g$ if one wants to
interpret these states as elementary particles \cite{fe/po/te}.  Of
course, that there is no horizon is disturbing but as it was shown in
\cite{ho/ma} even timelike singularities can be completely
non-singular when probed with quantum particles. In analogy to the
hydrogen atom this means that a classical singular theory becomes
nonsingular quantum mechanically. By the way, singularities play a
useful role - they can enable a stable ground state \cite{ho/my}.

\vs3

\noindent
{\em 3. The cases: $\a\neq0$, $\b\neq0$ , $\g=0$ or $\a\neq0$,
$\b\neq0$, $\g\neq0$.} Let us summarize both cases here. The first one
corresponds to two singular points along the line $y=z=0$ at $x=\pm
\sqrt{\a^2 + \b^2}$. In the second case the harmonic functions have no
poles at all. But nevertheless they have zeros and thus the metric has
singularities. This is interesting because in this case we have no
weak coupling region (dilaton $\f \rightarrow -\infty$ in (\ref{200}))
but only a strong coupling region (corresponding to the zeros). Since
our theory is not well defined there let us ignore this case
further. Perhaps it is worthwhile to look on these case a little bit
more carefully, but we want to leave this question for future
investigations.

\section{The S--duality and H--monopoles}

In this section we are going to discuss the strong-weak coupling (S-)
duality.  This symmetry is a string version of the old Montonen--Olive
conjecture that the theory is invariant under replacing the electric
with the magnetic charges and simultaneously the inversion of the
coupling constant. This symmetry has not yet been proved, but
nevertheless there are a lot of good reason to belief that this is
really a symmetry of the theory (for a nice short review see
\cite{sc}). The reason for the difficulties is that one knows the
theory only perturbatively (i.e.\ for weak couplings) and this
symmetry has a non-perturbative nature. It mixes the different orders
of the perturbative expansion, e.g.\ the weak with the strong coupling
region.

\vs5

The starting point is to transform the model into the Einstein
frame. This frame is defined by a conformal rescaling of the metric
in order to decouple the dilaton from the curvature in
(\ref{010}). Then one goes on-shell with the torsion, i.e.\ one
replaces the torsion by the axion $a$. Thus we write
\be410
G_{\m\n} \rightarrow G^E_{\m\n} = e^{-2\f} G_{\m\n} \quad , \quad
H_{\m\n\lambda} = - \sqrt{\det G^E} \, e^{4 \f} \,\e_{\m\n\lambda\s} \,
\p^{\s} \,a
\ee
where $G^E_{\m\n}$ is now the Einstein metric. As a next step one
combines the dilaton and the axion to one complex scalar field
\be420
\lambda = a + i e^{-2 \phi} \ .
\ee
Now, after these transformations in the 4D theory (coming from
the dimensional reduction of (\ref{010})) one finds that the equations
of motion are invariant under \cite{sh/tr/wi}, \cite{se4}
\be430
\ba{l}
\lambda \rightarrow \frac{a\, \lambda + b}{c\, \lambda +d} \quad , \quad
\mbox{with : } \quad ad-bc=1 \quad , \quad a,b,c,d \in \bf{R} \\
F_{\m\n}^{(m)} \longrightarrow (c \, Re \lambda  + d )\, F_{\m\n}^{(m)}
 \, + \, c \, Im \lambda \, (M L)_{mn} \tilde{F}_{\m\n}^{(n)}
\ea
\ee
where
\be440
\tilde{F}_{\m\n}^{(m)} = \frac{1}{2} (\det G^E)^{-\frac{1}{2}}
 G^E_{\m\m'} G^E_{\n\n'} \epsilon^{\m'\n'\lambda\rho}
 F_{\lambda\rho}^{(m)} \ .
\ee
The index $m$ numerates all gauge fields in (\ref{250}),
(\ref{260}) and the exact form of the matrices $M$ and $L$ can be
found in \cite{se4}.  All other fields, the (Einstein) metric and the
modulus field remains unchanged. The S-duality is therefore an
$SL(2,\bf{R})$ transformation in the complex scalar $\lambda$ and we
can generate new solutions which have in general both electric and
magnetic charges. Let us consider the special case where the
asymptotic values of the dilaton and the axion remains unchanged
($\f|_0=a|_0 = 0$), i.e.
\be450
\left(\ba{cc} a & b \\ c & d \ea \right) = \left( \ba{cc} \cos \g & \sin \g \\
      -\sin \g & \cos \g \ea \right) \in SO(2)
\ee
which means for the field strength
\be460
F_{\m\n}^{(m)} \rightarrow (-\sin \g \  a  + \cos  \g ) \,F_{\m\n}^{(m)}
 - \, \sin \g  \ e^{-2 \f} \, (M L)_{mn} \tilde{F}_{\m\n}^{(n)} \ .
\ee
{}From the asymptotic values we can get the new charges ($F_{0r}\propto
\frac{Q_e}{r^2}$ and $\tilde{F}_{0r} \propto \frac{Q_m}{r^2}$). For
the electric charges only the first part is relevant whereas the
magnetic charges are determined by the second part. Inserting the
asymptotic value for $(ML)^{(\pm)} \propto \pm {\bf 1}$ we find
\be470
\vec{Q}_{e}^{(\pm)} = \cos \g  \, \vec{Q}_{e}^{(\pm)old} \quad , \quad
\vec{Q}_{m}^{(\pm)} = \mp \sin \g \,  \vec{Q}_{e}^{(\pm)old}
\ee
or the new electric and magnetic charges are correlated by
\be480
\vec{Q}_{m}^{(\pm)} = \mp \tan \g  \vec{Q}_{e}^{(\pm)} \ .
\ee
The mass and the angular momentum in (\ref{381}) as defined via
the Einstein metric $G^E_{\m\n}$ remain fix. Again, we can express
the parameter $r_{\pm}$ of our solution by the new charges
\be481
r_{\pm} = \sqrt{2} \left( \, \sqrt{|\vec{Q}^{(+)}_e|^2 + |\vec{Q}^{(+)}_m|^2}
\pm  \sqrt{|\vec{Q}^{(-)}_e|^2 + |\vec{Q}^{(-)}_m|^2}\, \right)
\ee

\vs5

\noindent
{\bf The case $\g = \frac{\pi}{2}$ or $\lambda \rightarrow -
\frac{1}{\lambda}$.}  This transformation together with the trivial
discrete transformation $a \rightarrow a +1$ generate the discrete
subgroup $SL(2,\bf{Z})$.  This subgroup is expected to be a symmetry
of the whole quantum theory (by instanton effects the $SL(2,\bf{R})$
breaks down to this subgroup \cite{sh/tr/wi}). In this case our pure
electrically charged solution becomes a pure magnetically charged.
The transformed axion and dilaton are given by
\be490
a' = - \frac{a}{a^2 + e^{-4 \f}} \quad , \quad e^{-2\f'} =
\frac{e^{-2 \f}}{ a^2 + e^{-4 \f}}
\ee
where $a$ and $\f$ are the old axion and dilaton. We see that
for vanishing axion the dilaton changes the sign which corresponds
to the change from the weak to the strong coupling or vice versa.
Although the metric in the Einstein frame does not
change during the S-duality the metric in string frame
does (due to the new dilaton field)
\be500
ds'^2 = e^{2 \f'} ds_E^2 = (1 + a^2\, e^{4 \f}) \left[ (dt + \o_i dx^i)^2 -
e^{-4 \f} d\vec{x}^2 \right]
\ee
where $\f$ and $\o_i$ are defined in (\ref{340}) or (\ref{380}).
Let us discuss special cases.

\vs3

\noindent
{\em 1. S-self-duality.} Obviously, there is the special case if
\be510
a^2 +  \, e^{-4\f} = 1  \ .
\ee
In this case the dilaton remains unchanged and the axion changes only
the sign and because the effective action depends only quadratically
on the torsion this does not change the equations of motion. Only the
gauge fields get
a nontrivial transformation.  Let us call this case S-self-dual.
Inserting the dilaton and axion for the Taub-NUT case (\ref{310}),
(\ref{340}) we obtain the restriction
\be520
r_- + r_+ =0  \quad ,\quad r_+ r_- + 4 n^2 =0 \quad \leftrightarrow
\quad m=-\tilde{m} \quad , \quad n^2 = \tilde{m}^2 + |q|^2
\ee
i.e.\ the theory is S-self-dual iff it is T-self-dual (coming from
(\ref{130})) and the axion charge $n$ is correlated with the electric
charges ($\tilde{m}$ and $q_r$). Looking on the rotating black hole
configuration (\ref{361}) we find that asymptotically $a^2 \sim
r^{-4}$ but $e^{-4\f} \sim r^{-2}$ and therefore the rotating black
hole does not allow the S-self-dual configuration. By the way, the
different asymptotic behaviour of the axion is the reason why the
black hole configuration is asymptotically flat but the Taub-NUT
solution not. To clarify this case a little bit more let us consider
the gauge fields.  In the rotating black hole case we have seen that
our solution has only electric charges. In the Taub-NUT case we have
no asymptotically flat region and we cannot define charges as usual.
But looking on the asymptotic behavior of the gauge fields we can
define an electric and magnetic charge analogue. To simplify
let us assume that $\o_r=q_r=0$ and consider the gauge field
$A_{\m}^{\ m}$ (\ref{250})
\be521
A_{\m}^{\ m} = - \frac{1}{2} \, \frac{1}{1 - \frac{2 \tilde{m}}{r}}\,
\{ 1 , \o_i \}\, l^m
  \quad \mbox{with:} \quad  \o_i \,dx^i = 2 n \cos \theta \, d\f \ .
\ee
If we now look on the asymptotic behavior of the corresponding
field strength we find
\be522
F_{0r}^{\ m} \, \propto \, \frac{\tilde{m}}{r^2} \,l^m \qquad , \qquad
\tilde{F}_{0r}^{\ m} \propto  \frac{ n}{r^2} \, l^m \ .
\ee
Thus $\tilde{m}$ and $n$ play the role of an electric and magnetic
charge and the Taub-NUT solution carries both.  Furthermore in
self-dual case (\ref{520}) both charges coincide up to a possible
sign. If we add the other electric charges, i.e.\ $q_r \neq 0$ the
magnetic charge does not change but the electric charge gets
additional terms. The condition (\ref{520}) then simply means that the
length of the electric and magnetic charge vector is equal.  Finally
for the complex scalar function $\lambda$ we find
\be530
|\lambda|^2 = |a + i e^{-2\f}|^2 = (a^2 + e^{-4 \f})^2 =1 \ ,
\ee
i.e.\ for the self-dual case $\lambda$ is a unit vector. In this
case we do not change from weak to strong region or vice versa. Instead,
we remain in one region (the change of the dilaton is compensated by the
axion).

\vs3

\noindent
{\em 2.\ Vanishing axion ($a=0$).} In this case $\o_i$ vanish as well
($\p_i a = \epsilon_{ijk} \p_{j} \o_k$) and therewith the angular
momentum or Taub-NUT parameter. So, both solutions the rotating black
hole and the Taub-NUT coincides.  This case is interesting since here
the weak coupling region is directly transformed into a strong
coupling region ($\f \rightarrow -\f$) and therefore it is just the
opposite to the former case.

Setting the axion to zero we get an extreme magnetic black hole
coupling to 7 gauge fields
\be540
ds^2 = dt^2 - (1+\frac{r_+}{r})(1+\frac{r_-}{r}) d\vec{x}^2 \quad ,
\quad e^{2\f'} = \sqrt{(1+\frac{r_+}{r})(1+\frac{r_-}{r})}
\ee
where $r_{\pm}$ are given by (\ref{350}) and for the magnetic charges
one obtains
\be550
Q_m^{(+)r} = - \frac{1}{\sqrt{2}}\,(m+\tilde{m})\,l^r \quad , \quad
Q_m^{(-)r} = - \sqrt{2} \, (q^r + \frac{1}{2}(m-\tilde{m})\,l^r) \  .
\ee
For a flat internal space $r_+ = r_-$ ($q^r=0$ and $m=\tilde{m}$) and
one gets the standard extreme magnetic black hole coupling
to one gauge field \cite{ga/ho/st}. For the T-self-dual case ($m= -
\tilde{m}$) again the black hole becomes massless and neutral for the
gravi-photon ($Q_m^{(+)r}=0$).

Another limit yields the H-monopole solution. Here we have to set
$r_-=0$, i.e.\ $m \,\tilde{m} = |q|^2$. In this case both magnetic
charges in (\ref{481}) compensate each other $|Q_{m}^{(+)}| =
|Q_{m}^{(-)}|$ (note that for the $SL(2, {\bf Z})$ transformations we
have only magnetic charges). Therefore an H-monopole is nothing else
as an extreme magnetic black hole with a balance between the magnetic
charges coming from the matter and from the graviton sector. As long
as $q^r, \tilde{m} \neq 0$ this monopole couples to more gauge
fields. To get the standard H-monopole \cite{du/kh} coupling only to
one gauge field we set
\be560
m=q^r=0  \qquad (K=1)
\ee
and find
\be561
\ba{ccc}
ds^2 = dt^2 - (1+\frac{r_+}{r})d\vec{x}^2  & , &
 e^{2\f'} = \sqrt{(1+\frac{r_+}{r})} \\
e^{-2 \s} = 1 + \frac{r_+}{r} & , & F_{ij} = \frac{1}{2}
\epsilon_{ijk}\, \p_k (1 + \frac{r_+}{r})
\ea
\ee
where the gauge field is the S-dual of (\ref{250}). The gauge field
coming from the antisymmetric tensor (\ref{260}) is trivial in this
case.  This monopole solution can be obtained by a dimensional
reduction of the neutral fivebrane solution \cite{du/lu}. In this case
we have no gauge fields in the higher dimensional theory and the
magnetic charge comes from the axion charge (torsion). The
interpretation of this solution is that the fivebrane in 10 dimension
wraps around five of the six compactified dimensions.  This
H-monopole is a solution of the Kaluza--Klein (KK) theory as
well. Namely, instead of setting $m=0$ we can set $\tilde{m}=0$. This
corresponds to $F=1$ and thus a vanishing 10D dilaton ($\hat{\f}$, see
(\ref{090})). Also, the 10D antisymmetric tensor (\ref{140}) is
constant ($\o_i = \o_m = 1$) and therefore the torsion vanish too. So,
the 10D action is given only by the Einstein-Hilbert term.  The
corresponding electric charged solution (\ref{306}) is the S-dual to
this case. Both solutions confirm the statement that for a diagonal
internal metric the most general static KK black hole can have at
most one electric and one magnetic gauge field \cite{cv/yo}.

Interestingly all these magnetic solution are not yet exact in the
$\a'$ expansion. One can, however, promote these solutions to exact
ones by embedding the 10D spin connections into a non-abelian gauge
group. For the extreme magnetic black hole this has been done in
\cite{ka/or2} and for the 5-brane solution see e.g.\ in
\cite{du/lu}. This embedding is necessary to remove anomaly related
higher order terms in $\a'$. On the other side the electric charged
solutions are already exact without this embedding. The situation is
not very clear. If we start from the exact magnetic solution including
the non-abelian gauge field, there is no electric analogue. In
general it is impossible to embed the spin connection in a non-abelian
gauge field. Only for $F=K=1$ the holonomy group is compact (see
\cite{ho/ts1}). We have to postpone this question until the S-duality
transformation of non-abelian gauge fields is better understood.

Despite this shortcoming it is remarkable that the S-duality relates
solutions which are exact in the $\a'$ expansion (after this
embedding). The point is that this relation between two exact
backgrounds is independent of the ``starting dimension''.  As long as
we start with a 10D theory and reduce it down to 4 dimenions we get a
theory with $N=4$ and one expects that there are no quantum
corrections and the S-duality maps in fact exact solutions to exact
solutions (see e.g.\ \cite{se4}). But this relation is also valid for
a 6D theory reduced to 4 dimensions yielding $N=2$. There, one would
expect corrections and the S-duality mixes the orders, i.e.~in general
a solution in the lowest order is not expected to transform again in a
solution of the lowest order. In other words, one would expect that
corrections of the higher genus or even non-perturbative terms
contribute in the transformed theory to the lowest order (i.e.\ lowest
genus).  But for both types of solution one cannot see such
corrections. Even more, above we have seen that our electric solution
is near the singularities asymptotically free. This is a hint that
higher genus contribution and non-perturbative corrections are there
under control. On the other side the magnetic solutions are there in
the strong coupling region, i.e.\ one could expect a strong influence
of higher genus and non-perturbative contributions.  But no such
corrections can be seen.  Although one knows that the string fivebrane
is in the strong coupling region effectively a SU(2) WZW model
\cite{du/lu} ($R \times S^3$ throat limit) and therefore an exact
conformal field theory, this is not the case for the H-monopoles. To
get this monopole solution one assumes that the fields of the
fivebrane solution \cite{du/kh} depend only on three coordinates and
thus one destroys the $S_3$ symmetry. This procedure here suggests
that also the H-monopole remains exact in the strong coupling region
(as long as S-duality is real a symmetry).  Furthermore, if we go the
other direction, since the H-monopole solution is an exact model away
from the singularity we can speculate that for our original model
(\ref{010}), even away from the weak coupling region, further
corrections (no higher genus nor non-perturbative) are negligible.
\bigskip

\noindent
{\bf Identification of the string states.} Finally, let us try
to identify a subset of our solutions in the elementary string
spectrum. As usual let us consider the mass formula for string
excitations in the Neveu-Schwarz (NS) sector which
is given by
\be562
M^2 = \frac{1}{2} \left( \, |\vec{Q}_e^{(+)}|^2 + 2(N_R -\frac{1}{2})\,
\right) = \frac{1}{2} \left(\, |\vec{Q}_e^{(-)}|^2 + 2(N_L - 1 )\, \right)
\ee
where $\vec{Q}_e^{(\pm)}$ is the electric charge vector (internal
momentum contributions) which form an even selfdual lattice and
$N_{R/L}$ are the oscillator contributions. We know that the right moving
sector of our solution saturates the Bogomol'nyi bound (\ref{391})
which means
\be563
N_R = \frac{1}{2} \ .
\ee
This is the lowest possible value required by GSO projection. Then
using (\ref{382}) we find for the left moving sector
\be564
N_L - 1 = \frac{1}{2} \left(\, |\vec{Q}_e^{(+)}|^2 - |\vec{Q}_e^{(-)}|^2
\, \right) = m \, \tilde{m} - |q|^2 \ .
\ee
Of course, because the charges enter the formular as internal momenta
this equation describes first of all electrically charged solutions,
i.e.\ in our case electric black holes. Thus they are interpreted
as elementary string states. On the other side the magnetic solutions
appearing as the S-dual of the electric solutions are then identified
as soliton excitations of the theory. Let us now discuss special cases
of our solution.
\smallskip

\noindent
{\em 1. T-self-dual case.} Here we have $m=-\tilde{m}$ and the black
hole solutions (electric or magnetic) become massless.  Inserting this
into the mass formula yields $N_L -1 = - (\tilde{m}^2 +
|q|^2)$. Therefore, our massless black holes can be identified as
string excitation with $N_L = 0$. In principle the same is true for
the S-self-dual case. But there the charges are not well defined (the
Taub-NUT geometry is asymptotically not flat). Note, that for this
case it is crucial to have at least two gauge fields\footnote{See
(\ref{302}), as long as $F K \neq 1$ we have always at least two gauge
fields.}.  This is maybe the reason why Duff and Rahmfeld \cite{du/ra}
were unable to identify a state with $N_L=0$. Sen \cite{se4} could
identify such a state as BPS monopoles, i.e.\ a monopole solution
where the 10D theory contains a non-abelian gauge field.
\smallskip

\noindent
{\em 2. H-monopoles, Kaluza-Klein case.} Our generalization of
the H-monopoles are defined by $m \tilde{m} = |q|^2$ and therefore
we get $N_L=1$. This coincides with the results for the standard
H-monopole with only one gauge field ($|q|=0$) \cite{du/ra,se4}.
As we have already seen, this solution is nothing else as
the extreme electric (\ref{306}) or extreme magnetic (\ref{561})
KK black hole.
\smallskip

\noindent
{\em 3. Extreme dilaton black holes, rotating black holes, ...} All
other cases have $N_L>1$. Special examples with
only one gauge field and flat internal space ($|q|=0$, $m=\tilde{m}$)
\cite{ka/ka/or/to} are, e.g., the rotating dilaton black hole solution
(with a nacked singularity) or the extreme dilaton black holes.

\section{Conclusion}
In this paper we have started with a general model which allows a null
Killing vector and which has unbroken spacetime supersymmetries.  This
model also known as chiral null model, is the generalization of the
gravitational wave and fundamental string background. As discussed in
section two this model possesses a chiral symmetry on the world sheet and
is exact in the $\a'$ expansion. In addition, there are points of
explicit T-self-duality. Regarding this model as the bosonic part of a
$D=10$, $N=1$ superstring background we have reduced
this model to 4 dimensions.  As result we got a stationary IWP solution
(\ref{300}) which couples to 7 gauge fields (one gravi-photon and 6
matter gauge fields) and 4 scalars (dilaton, axion and two moduli).
This solution is completely determined by ten harmonic
functions which can have different kind of singularities yielding
different geometries. Assuming that we have a point like singularity
we get the Taub-NUT geometry (\ref{340}) whereas a ring singularity
yields an electric Kerr black hole (\ref{380}). In addition to these
both cases the harmonic function can be non-singular or can have
singularities in two points. Whether these cases have a reasonable
space time interpretation remains unclear. As expected, in the
T-self-dual point the black hole becomes massless and the charge
correlated to the gravi-photon vanishes.  Furthermore, we showed that
the gravi-photon state saturate a Bogomol'nyi bound.  Unfortunately,
this black hole solution has a naked singularity as long as the
angular momentum does not vanish. On the other side, there are
arguments to interpret these states with the naked ring singularity
as elementary closed strings.

In the last section we have investigated the S-duality for this
solution.  After some general remarks we have discussed the $SL(2,{\bf
Z})$ transformation $\lambda \rightarrow - \lambda^{-1}$.  There are
two special cases. One is the S-self-dual limit where the dilaton and
the axion remains unchanged the S-duality transformed not the weak
into the strong coupling region.  This case is only possible for the
Taub-NUT geometry. Secondly, we have shown that for vanishing axion
the magnetic solution is an extreme magnetic black hole (\ref{540}) or
an H-monopole (\ref{561}). Both solutions couple to more gauge fields
and in the limit of only one gauge field we obtained the standard
expressions. It turned out that an H-monopole is nothing else as an
extreme magnetic black hole with a balance between the magnetic
charges coming from the matter and from the graviton sector.
Interestingly, both solutions are not exact in the $\a'$ expansion.
There are anomaly related $\a'$ corrections, but one can promote these
solutions to exact ones by embedding the torsionfull 10D spin
connections into a non-abelian gauge group.  Such an embedding can
only be understood in the framework of heterotic string model with a
non-abelian gauge field in 10 dimensions.  On the other hand the
electric solutions are exact already without this embedding and can
be regarded as a pure bosonic string model.  Maybe a better
understanding of the S-duality for non-abelian gauge fields can yield
a better understanding of this phenomena.

In the last subsection we have tried to identify our solutions as
elementary string or soliton excitations of the theory. We found,
that the T-self-dual solution corresponds to states with $N_R=\frac{1}{2}$,
$N_L=0$, the H-monopole class to $N_R=\frac{1}{2}$, $N_L=1$ and all
other solutions including the extreme dilaton and rotating  black holes
to $N_R=\frac{1}{2}$, $N_L>0$.

In our consideration it remains an open problem why the massless
black holes or string excitations are at rest and does not move with
the speed of light.

\vspace{10mm}

\noindent {\large\bf Acknowledgments}
\vspace{3mm}
\newline
\noindent
I would like to thank H. \ Dorn for reading the manuscript and
suggesting improvements. Also I am grateful to R.\ Kallosh for
usefull comments and A.\ Tseytlin for several discussions.

\newpage
\renewcommand{\arraystretch}{1.0}


\begin{thebibliography}{aaa}
\bibitem{ts2}
A.A. Tseytlin, ``Exact solutions of closed string theory'',
Imperial/TP/94-95/28 (hep-th/9505052)
\bibitem{ho/ts1}
G.T. Horowitz and A.A. Tseytlin, ``A new class of exact solutions in string
theory'', \PRD51 (1995) 2896;
G.T. Horowitz and A.A. Tseytlin, ``On exact solutions and singularities in
string theory'', \PRD50 (1994) 5204;
\bibitem{ho/ste}
G. Horowitz and A. Steif, \PRL64 (1990) 260, \PRD42 (1990) 1950 ;\\
R.E. Rudd, \NP352 (1991) 489;\\
A.A. Tseytlin, \NP390 (1993) 153;\\
R. G\"uven, \PL191 (1987) 275.
\bibitem{kl/ts}
C. Klimcik, A.A. Tseytlin, \NP424 (1994) 71.
\bibitem{da/gi/ha/ru}
A. Dabholkar, G. Gibbons, J. Harvey, F. Ruiz Ruiz, \NP340 (1990) 33.
\bibitem{be/ka/or}
E. Bergshoeff, R. Kallosh and T. Ort\'{\i}n, ``Black--Hole--Wave duality
in string theory'', \PRD50 (1994) 5188.
\bibitem{ka}
E. Bergshoeff, R. Kallosh and T. Ort\'{\i}n, \PRD47 (1993) 5444;\\
E. Bergshoeff, I. Entrop and R. Kallosh, \PRD49 (1994) 6663; \\
E. Kiritsis, C. Kounnas, D. L\"ust, \PL331 (1994) 321.
\bibitem{ka/li}
R. Kallosh, A. Linde, ``Exact supersymmetric massive and massless white
holes'', preprint SU-ITP-9514 (hep-th/9507022).
\bibitem{ka/ka/or/to}
R. Kallosh, D. Kastor, T. Ort\'{\i}n and T. Torma, ``Supersymmetry and
stationary solutions in dilaton--axion gravity'', \PRD50 (1994) 6374.
\bibitem{be}
K. Behrndt, ``The 10-D chiral null model and the relation to 4-D
 string solutions'', \PL348 (1995) 395.
\bibitem{na/wi}
C.R. Nappi, E. Witten, ``A WZW model based on a nonsemisimple group'',
\PRL71 (1993) 3751 (hep-th/9310112).
\bibitem{wa}
D. Waldram, \PRD47 (1993) 2528; J.P. Gauntlett, J.A. Harvey,
M.M. Robinson and D. Waldram, \NP411 (1994) 461.
\bibitem{ma/sc}
J. Maharana and J.H. Schwarz, ``Noncompact symmetries in string theory',
\NP390 (1993) 3.
\bibitem{du}
M.J. Duff, ``Kaluza-Klein theory in perspective'', preprint
NI-94-015 (hep-th/9410046).
\bibitem{ch}
A.H. Chamseddine, ``N=4 supergravity coupled to N=4 matter
and hidden symmetry'', \NP185 (1981) 403.
\bibitem{ru/ts}
J.R. Russo, A.A. Tseytlin, ``Constant magnetic field in closed string
theory: an exact solvable model'', preprint CERN--TH.7494/94
(hep-th/9411099).
\bibitem{is/wi/pe}
Z. Perj{\'e}s, \PRL27 (1971) 1668; \\
W. Israel and G.A. Wilson, {\em J. Math. Phys.} {\bf 13} (1972) 865.
\bibitem{to}
K.P. Tod, \PL121 (1983) 241.
\bibitem{gi}
G. Gibbons, \NP207 (1982) 337\\
G. Gibbons and M.J. Perry, \NP248 (1984) 629\\
G. Gibbons and K. Maeda, \NP298 (1988) 741.
\bibitem{ho/st}
G. Horowitz and A. Strominger, ``Black strings and p-branes'',
  \NP360 (1991) 197.
\bibitem{ts}
A.A. Tseytlin, ``Black holes and exact solutions in string theory'',
to be published in the proceedings of {\it International School
of Astrophysics}(D. Chalonge): 3rd Course: Current Topics
in Astrofundamental Physics, Erice, Italy, 4-16 Sep 1994.
\bibitem{jo/my}
C.V. Johnson and R.C. Myers, ``Taub--NUT dyons in heterotic string theory,
\PRD 50 (1994) 6512.
\bibitem{mi}
C. Missner, {\em J. Math. Phys.} {\bf 4} (1963) 924.
\bibitem{ka2}
R. Kallosh, ``Duality symmetric quantization of superstring'', preprint
SU-ITP-95-12 (hep-th/9506113).
\bibitem{se2}
A. Sen, \MPLA8 (1993) 2023 (hep-th/9303057).
\bibitem{se}
A. Sen, ``Black hole solutions in heterotic string theory on a torus'',
\NP440 (1995) 421 (hep-th/9411187).
\bibitem{ni}
H. Nishino, ``Stationary axisymmetric black holes, $N=2$
superstring, and self-dual gauge or gravity fields'',
preprint UMDEPP 95-111 (hep-th/9504142).
\bibitem{se3}
A. Sen, ``Extremal black holes and elementary string states'',
preprint TIFR-TH-95-19 (hep-th/9504147).
\bibitem{ru/su}
J. Russo and L. Susskind, ``Asymptotic level density in heterotic string
theory and rotating black holes'', preprint UTTG--9--94 (hep-th/9405117).
\bibitem{fe/po/te}
S. Ferrara, M. Porrati and V.L. Telegdi, \PRD46 (1992) 3529.
\bibitem{ho/ma}
G.T. Horowitz and D. Marolf, ``Quantum probes of spacetime
singularities'', preprint UCSBTH-95-5 (gr-qc/9504028).
\bibitem{ho/my}
G. Horowitz and R. Myers, ``The value of singularities'',
preprint UCSBTH-95-6, McGill/95-20 (gr-qc/9503062).
\bibitem{sc}
J.H. Schwarz, ``String theory symmetry'', CALT-68-1984
(hep-th/9503127)
\bibitem{sh/tr/wi}
A. Font, L. Ibanez, D. L\"ust and F. Quevedo, \PL249 (1990) 35; \\
A. Shapere, S. Trevedi and F. Wilczek, \MPLA6 (1991) 2677.
\bibitem{se4}
A. Sen, ``Strong--weak couplings duality in four dimensional string
theory'', \IMP9 (1994) 3707 (hep-th/9402002).
\bibitem{ga/ho/st}
D. Garfinkle, G.T. Horowitz and A. Strominger, \PRD43 (1991) 3140.
\bibitem{du/kh}
R. Khuri, \NP387 (1992) 315;
M.J. Duff and R.R. Khuri, ``Four-dimensional string/string duality'',
\NP411 (1994) 473;
M.J. Duff, R.R. Khuri, R. Minasian and J. Rahmfeld, \NP418 (1994) 195;
J. Gauntlett, J. Harvey and J. Liu, \NP409 (1993) 363.
\bibitem{du/lu}
M.J. Duff, J.X. Lu, \NP354 (1991) 141; \\
C.G. Callan, J.A. Harvey, A. Strominger, \NP359 (1991) 611.
\bibitem{cv/yo}
M. Cveti\v{c} and D. Youm, ``Static four-dimensional abelian black holes
in Kaluza--Klein theory'', preprint UPR-645-T (hep-th/9502099)
\bibitem{ka/or2}
R. Kallosh and T. Ort\'{\i}n, ``Exact $SU(2) \times U(1)$ stringy black
hole'', \PRD50 (1994) 7123 (hep-th/9409060).
\bibitem{du/ra}
M.J. Duff, J. Rahmfeld, ``Massive string states as extreme black
holes'', \PL345 (1995) 441 (hep-th/9406105).
\end{thebibliography}
\end{document}